\documentclass[twocolumn]{revtex4}

\usepackage{graphicx}
\usepackage{setspace}
\usepackage{multirow}

\setcitestyle{square}

\begin{document}

\title{Vibrational branching ratios and hyperfine structure of $^{11}$BH and its suitability for laser cooling}
\author{R. J. Hendricks, D. A. Holland, S. Truppe, B. E. Sauer and M. R. Tarbutt}
\email{m.tarbutt@imperial.ac.uk}
\address{Centre for Cold Matter, Blackett Laboratory, Imperial College London, Prince Consort Road, London SW72AZ, UK}

\begin{abstract}
The simple structure of the BH molecule makes it an excellent candidate for direct laser cooling. We measure the branching ratios for the decay of the ${\rm A}^{1}\Pi (v'{=}0)$ state to vibrational levels of the ground state, ${\rm X}^{1}\Sigma^{+}$, and find that they are exceedingly favourable for laser cooling. We verify that the branching ratio for the spin-forbidden transition to the intermediate ${\rm a}^{3}\Pi$ state is inconsequentially small. We measure the frequency of the lowest rotational transition of the X state, and the hyperfine structure in the relevant levels of both the X and A states, and determine the nuclear electric quadrupole and magnetic dipole coupling constants. Our results show that, with a relatively simple laser cooling scheme, a Zeeman slower and magneto-optical trap can be used to cool, slow and trap BH molecules.

%\tiny
% \keyFont{ \section{Keywords:} Vibrational branching ratios, Frank-Condon factors, hyperfine structure, cooling molecules, boron hydride, laser spectroscopy, mm-wave spectroscopy }

\end{abstract}

\maketitle

\section{Introduction}

Laser cooling has been applied with great success to a wide variety of atomic species, leading to huge advances in many fields including metrology, sensing, interferometry, tests of fundamental physics, studies of ultracold collisions and studies of quantum degenerate gases. There is currently great interest in extending the laser cooling method to molecules, motivated by a similarly rich host of applications in fundamental physics and quantum chemistry \cite{Carr09}. Direct laser cooling has recently been demonstrated for three molecular species, SrF~\cite{Shuman10, Barry12}, YO~\cite{Hummon2013} and CaF~\cite{Zhelyazkova13}, and laser cooling of YbF is also being explored~\cite{Tarbutt2013a}. For SrF, a magneto-optical trap has recently been demonstrated \cite{Barry14}. For all these molecules, the laser cooling transition is between the ground $^{2}\Sigma^{+}$ state and an electronically excited $^{2}\Pi_{1/2}$ state. For laser cooling to be feasible, the molecule must have a short-lived excited state that decays with very high probability to just one or a few vibrational levels of the ground state, at wavelengths that are easily produced with current laser technology. There should be no accessible intermediate state, and the molecule should have a sufficiently simple rotational and hyperfine structure. Fortunately, there is quite an extensive list of candidate molecules \cite{diRosa04}, though in many cases new data is needed to assess their suitability.

\begin{figure*}[tb]
\centering
\includegraphics[width=0.9\textwidth]{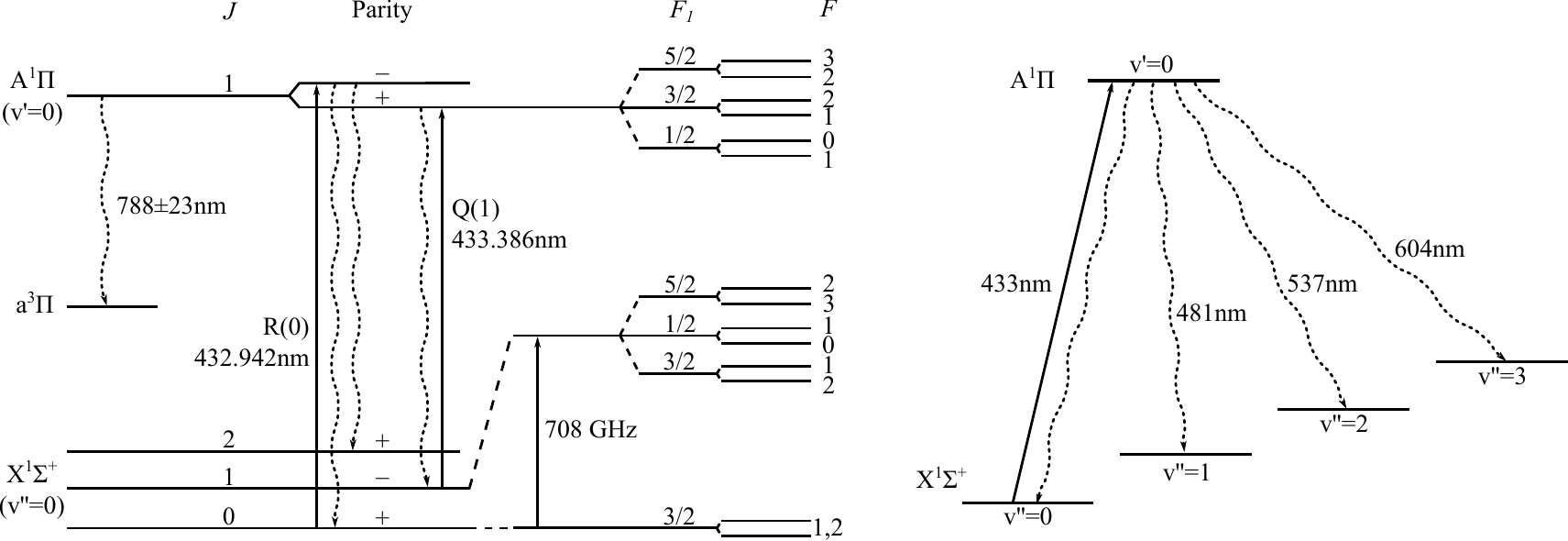}
\caption{Structure of the lowest lying ro-vibrational levels of electronic states in $^{11}$BH relevant for laser cooling.  The cooling transition is the indicated Q(1) line. Wavy lines show allowed decay paths. The spin-forbidden transition to the intermediate ${\rm a}^3\Pi$ state is strongly suppressed.}
\label{fig:levelstruct}
\end{figure*}

As we discuss here, molecules with a $^{1}\Sigma$ ground state and $^{1}\Pi$ excited state, such as BH, are particulary attractive candidates for laser cooling, though none have yet been cooled. Figure \ref{fig:levelstruct} shows the relevant energy levels of ${}^{11}$BH. In the ground state, ${\rm X}^{1}\Sigma^{+} (v''{=}0)$, there is a ladder of rotational states of alternating parity. In the electronically excited state, ${\rm A}^{1}\Pi (v'{=}0)$, each rotational state is a pair of opposite-parity levels split by the $\Lambda$-doubling interaction. The main transition of interest for laser cooling is the ${\rm A}^1\Pi (v^\prime{=}0, J^\prime{=}1) \longleftarrow {\rm X}^1\Sigma^{+} (v''{=}0, J''{=}1)$ electric dipole transition at 433nm~\cite{Johns67, Fernando91}, labelled Q(1) in figure \ref{fig:levelstruct}. The upper state lifetime is $127\pm10$\,ns~\cite{Rice89}, permitting rapid photon scattering as is desirable for laser cooling. Due to the selection rules for the change in parity and angular momentum in an electric dipole transition, the upper state decays exclusively on the Q(1) branch, always returning the molecule to $J''=1$. The same is true of all the other Q-branch lines: all are `rotationally closed'. This means that molecules in every rotational state are amenable to laser cooling, with the exception of the ground state which cannot be excited on a Q-line. The upper state can, of course, decay to other vibrational levels of X, but for BH the branching ratios for these other transitions are expected to be small. Interestingly, the ground state has a magnetic g-factor very close to zero, while the upper state has $g\simeq 1$. In a strong magnetic field a single Zeeman sublevel of the upper state can be excited by the laser, while the lower sub-levels remain unresolved. This is an ideal situation for Zeeman slowing, and is in contrast to molecules that have a ${}^{2}\Sigma$ ground state where the complexity of the Zeeman splitting renders Zeeman slowing unpalatable. Finally, in a ${}^{1}\Sigma$ state the hyperfine structure is likely to be smaller than the linewidth of the laser cooling transition, so that all hyperfine components are addressed without needing to apply sidebands to the lasers. Figure \ref{fig:levelstruct} shows the hyperfine structure of the X and A states of ${}^{11}$BH, which is discussed in more detail below.

In this paper we measure the key properties needed to determine a feasible scheme for laser cooling and Zeeman slowing of BH. We measure the branching ratios from the ${\rm A}(v'{=}0)$ state to the various vibrational states of X. We measure the frequency of the first rotational transition, and the hyperfine structure in both the X and A states. The molecule has a triplet state, $a^3\Pi$, lying between the A and X states, and decay to this state may be a limitation to laser cooling. We measure an upper limit to the branching ratio for this spin-forbidden transition.

%%%%%%%%%%%%%%%%%%%%%%%%%%%%%%%%%%%%%%%
%%%%%%%%%%%%%%%%%%%%%%%%%%%%%%%%%%%%%%%

\section{Methods}

\begin{figure}[tb]
	\centering
		\includegraphics[width=0.45\textwidth]{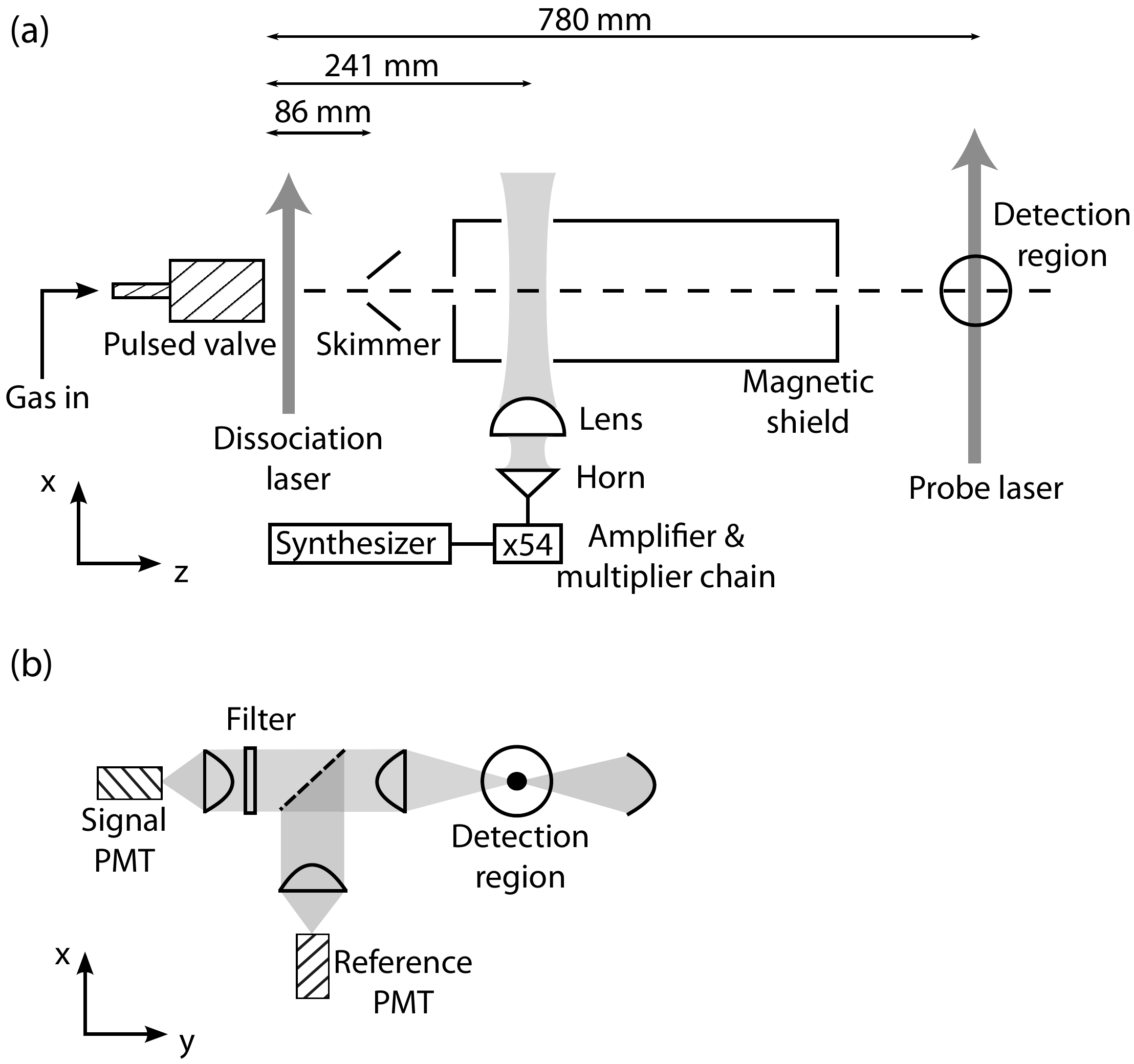}
		\caption{(a) Apparatus for producing and detecting a supersonic beam of BH molecules, and for driving transitions between rotational states. (b) Laser-induced fluorescence detection setup. Using a set of interference filters placed in the arm containing the signal PMT, the branching ratios to various vibrational states is determined. The reference PMT is used to account for fluctuations in molecular flux.  }
		\label{fig:setup1}
\end{figure}

A schematic of the experiment is shown in figure~\ref{fig:setup1}(a).  A supersonic beam of cold BH molecules is produced by photodissociation of a diborane ($\mathrm{B}_2 \mathrm{H}_6$) precursor, at 0.6\% concentration in argon, following previous methods for producing BH~\cite{Clark01} and CH \cite{Truppe(2)14}. At a pressure of 3.5\,bar the gaseous precursor feeds a solenoid valve with a 1\,mm orifice, which is briefly opened with a $160\,\mu$s pulse of current. The 193\,nm light from an excimer laser, with a pulse duration of 20\,ns and energy of 120\,mJ, is focussed onto the gas pulse exiting the valve causing dissociation to a variety of products including, by a two-photon process, the BH molecule~\cite{Harrison88}. The source operates with a repetition rate of 10\,Hz and the mean pressure in the source chamber is $10^{-4}$\,mbar. The molecules pass through a skimmer 86\,mm downstream from the valve nozzle into a chamber where the background pressure is $10^{-7}$\,mbar. The beam has a speed of 570\,m/s and a translational temperature of 0.4\,K.

About 85\% of the BH molecules are in the ground rotational state. Following the methods detailed in \cite{Truppe(1)14}, we drive the first rotational transition using millimetre-wave radiation at 708\,GHz. About 1\,$\mu$W of radiation at this frequency is produced by an amplifier-multiplier chain unit which generates the 54th harmonic of a frequency synthesizer, phase-locked to a 10\,MHz GPS reference. A diagonal horn antenna couples the millimetre-wave radiation into an approximately Gaussian beam, which is collimated by a 30\,mm focal length PTFE lens. This beam passes into the vacuum chamber through a PTFE window and crosses the molecular beam at 90$^{\circ}$, 155\,mm downstream from the skimmer.

The molecules travel a further 540\,mm to a laser-induced fluorescence detection region, where the frequency-doubled output from a Ti:sapphire laser excites the ${\rm A}^1 \Pi (v'{=}0) \longleftarrow {\rm X}^1 \Sigma^{+} (v''{=}0)$ electronic transition at 433\,nm. The laser has a linewidth of about 100\,kHz and is locked to an optical transfer cavity that is in turn locked to a He:Ne laser with a long-term stability of about 2\,MHz. We drive either the R(0) or Q(1) transition (figure \ref{fig:levelstruct}) to measure the population in the $J{=}0$ or $J{=}1$ components of the $X^1 \Sigma^{+}$ state. The probe laser beam has a power of 15\,mW, a waist of 1\,mm along the molecular beam axis and 5\,mm perpendicular to it, and crosses the molecular beam at right angles. The spectral distribution of the laser-induced fluorescence is measured using the apparatus shown in figure~\ref{fig:setup1}(b). The light is collimated by an aspheric condenser lens mounted close to the probe region, is split by a 50:50 non-polarising beam splitter, and then focussed onto two photo-multiplier tubes (PMTs), each operated in photon counting mode with a time resolution of 10\,$\mu$s. Interference filters with a bandwidth of 10\,nm are placed in one arm to select the fluorescence from individual vibrational branches of the transition. The other arm always monitors the unfiltered fluorescence to provide a reference signal proportional to the number of molecules in the beam.

%%%%%%%%%%%%%%%%%%%%%%%%%%%%%%%%%%%%%%%
%%%%%%%%%%%%%%%%%%%%%%%%%%%%%%%%%%%%%%%

\section{Results}

\subsection{Vibrational branching ratios}

To know which, and how many, laser wavelengths are needed for laser cooling, we need to know the branching ratios from ${\rm A} (v')$ to each of the vibrational states ${\rm X} (v'')$. These branching ratios are given by the ratio of Einstein coefficients $A_{v',v''}/(\sum_{v''}A_{v',v''})$. The A-coefficients are determined by the Franck-Condon factors, corrected for the dependence of the transition moment on the internuclear separation and the frequency dependence of the spontaneous emission rate. Being very light, BH has large vibrational frequencies that are a significant fraction of the dissociation energy, and so even the low-lying vibrational wavefunctions extend into regions where the molecular potential is significantly anharmonic. Theoretical vibrational branching ratios are therefore particularly sensitive to the exact form of the potential used in the calculation, and there is significant disagreement amongst the various calculations~\cite{Asthana71,L&S72,L&S83,Diercksen87}. For example, the predicted Franck-Condon factor for the decay to $v''=0$ varies from 67.48\% to 99.87\%.

\begin{figure}
	\centering
		\includegraphics[width=0.45\textwidth]{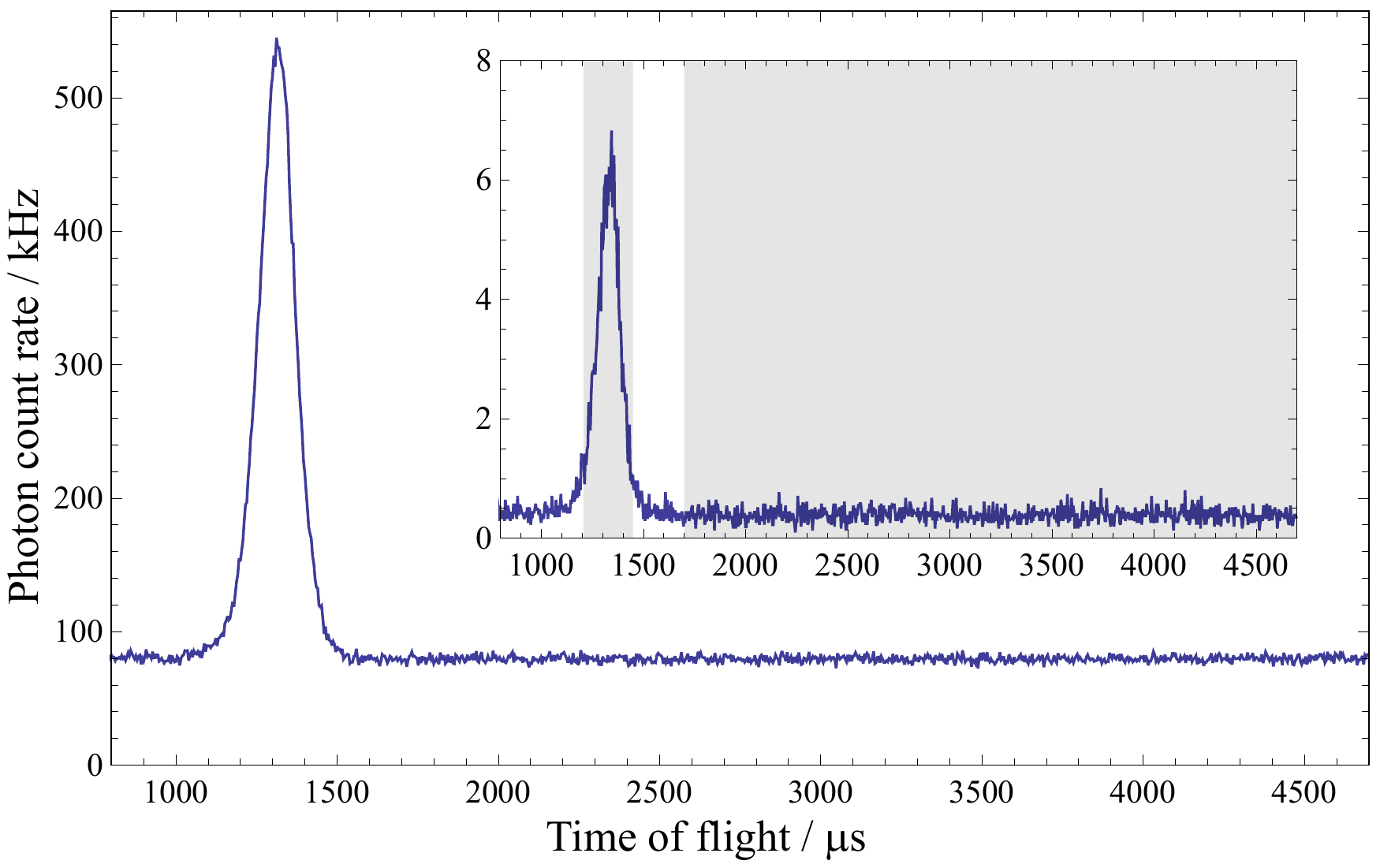}
		\caption{Time-of-flight profiles measured with two different filters in the signal arm of the detection setup, one isolating the decay to $v''=0$ (main plot) and the other to $v''=1$ (inset). The dissociation laser fires at $t=0$. The shaded regions in the inset show the time windows used to determine the number of signal and background photons.}
		\label{fig:TOF}
\end{figure}

Figure \ref{fig:TOF} shows time-of-flight profiles measured using two different interference filters, one isolating the decay to $v''=0$ and the other to $v''=1$. The molecules are excited on the Q(1) transition. We count the photons detected in the 230\,$\mu$s time window indicated in figure~\ref{fig:TOF}, corresponding to the period when there is a significant flux of molecules. The background due to laser scatter and ambient light is determined by counting the photons received in a 3000\,$\mu$s time window when there are no molecules present and dividing by the ratio of the two time periods. After subtracting this background, the signal contains two contributions, the laser-induced fluorescence we wish to measure, plus any additional molecular fluorescence which is not induced by the probe laser (e.g. due to long-lived states excited in the source). This background fluorescence is 0.04\% of the laser-induced fluorescence. We switch the probe laser beam on and off using a mechanical shutter so that alternate pulses record the fluorescence with and without the probe beam. The difference between these two, each with background subtracted, is the laser-induced fluorescence signal. We average such signals over several thousand molecular pulses for each filter in turn, and repeat multiple times using the filters in a random order. Variation in molecular flux is accounted for by dividing by the signal in the reference PMT, similarly processed.

The transitions to $v''>0$ are weak, and so it's particularly important to measure how much of the dominant 433\,nm fluorescence to $v''=0$ is transmitted by the filters used to isolate the $v''>0$ transitions. We therefore measure the transmittance of each of the filters at 433\,nm. The transmittance at other wavelengths is less critical and we use the manufacturer's data. We also measure the transmittances of the lenses and beamsplitters, and calibrate the relative response of the signal PMT using a lamp, grating spectrometer, and calibrated silicon photodiode.

The branching ratios obtained from the measured fluorescence yields, transmittances, and PMT response, are shown in Table \ref{tab:FCF}. The uncertainties given in the table include the statistical uncertainties of the measurements and all uncertainties arising from the calibration of the filters, optics and PMT. A previous measurement found $A_{01}/A_{00}=0.0051(7)$~\cite{Rice89}, which differs from our result by 4 standard deviations. None of the other branching ratios were measured previously. Table \ref{tab:FCF} also gives the branching ratios derived from theoretical calculations by Luh and Stwalley~\cite{L&S83}. These all agree with our measurements to an absolute accuracy of 0.004, and to within 2.5 standard deviations. Our experimental values are normalized so that their sum is unity, assuming that the branching ratios to all $v''>3$ are much smaller than those measured. From \cite{L&S83} we find the branching ratios to $v''>3$ to be 10$^{-7}$ or less, justifying this assumption. For $v^{\prime \prime}{=}0$, 1 and 2 the uncertainty is dominated by the uncertainties in the relative transmission of the detection optics and sensitivity of the PMT at the different wavelengths. Our $v^{\prime \prime}{=}3$ upper limit is limited by the statistical uncertainty in the measured difference between probe on and off.

The results in table~\ref{tab:FCF} are the branching ratios following excitation on the Q(1) transition. In the Born-Oppenheimer approximation the rotational state has no effect on the vibrational branching ratios, but at this level of precision, and for such a light molecule, it is worth investigating whether this approximation is sufficient. If the molecule is excited on the R(0) transition instead of Q(1), the final rotational states are $J=0$ and $J=2$, instead of $J=1$, as shown in figure~\ref{fig:levelstruct}. We repeated our measurements of the branching ratios, now exciting on the R(0) transition, and we find identical results to within the uncertainties given in the table.

\begin{table}[!t]
\begin{centering}
{\begin{tabular}{ cr@{.}l@{ }cc }\toprule
& \multicolumn{3}{c}{Experimental} & Theoretical \\
\hline
$v^{\prime \prime}{=}0$ &0&9863 & (19) & 0.99054\\
$v^{\prime \prime}{=}1$ &0&0128 & (18) & 0.00888\\
$v^{\prime \prime}{=}2$ &0&00093 & (15) & 0.00057\\
$v^{\prime \prime}{=}3$ &$<$0&00007 &  & $9.46 \times 10^{-6}$\\
\botrule
\end{tabular}}{}
\caption{Probabilities for decay from $A^1\Pi (v^\prime{=}0)$ to $X^1\Sigma^+ (v^{\prime \prime})$. The bracketed numbers are the $1\sigma$ uncertainties in the final digits. For $v^{\prime \prime}{=}3$ we give the 90\% confidence upper limit. Theoretical values are derived from~\cite{L&S83}.}
\label{tab:FCF}
\end{centering}
\end{table}

\subsection{The spin-forbidden ${\rm A}^1 \Pi \longrightarrow {\rm a}^3 \Pi$ transition}

Molecules that decay on the spin-forbidden ${\rm A}^1 \Pi \longrightarrow {\rm a}^3 \Pi$ transition will be lost from the laser cooling cycle, and so it is important to know the rate for this transition. There are no measurements or calculations of this rate, but we expect it to be of a similar magnitude to the rate for the ${\rm a}^3\Pi \longrightarrow X^1\Sigma^{+}$ transition, which is calculated to be less than 0.1\,s$^{-1}$~\cite{Pederson94}. The energy separation of the X and a states has been measured to within a few percent from the difference in their dissociation energies~\cite{Brazier96,Dagdigian97}, and this is in agreement with values obtained theoretically~\cite{Pople84,Pederson94}. From these, we expect the wavelength of the ${\rm A}^1\Pi\longrightarrow {\rm a}^3\Pi$ transition to be at 788$\pm$23nm. We use a broadband interference filter to isolate light in this wavelength range and search for fluorescence from this transition. We see none, and thus determine a 90\% confidence upper limit to the branching ratio for this decay route to be $3.4 \times 10^{-4}$, limited by the statistical uncertainty.

\subsection{Hyperfine structure in the excited state}

The hyperfine structure of the A state has not been measured or calculated previously, and it is important for identifying the best laser cooling scheme and for calculating the Zeeman effect of the excited state. To describe the hyperfine structure we use the Hamiltonian
\begin{equation}
H = a_\mathrm{B} (\mathbf{I}_{\mathrm{B}} \cdot \mathbf{L}) + a_\mathrm{H} (\mathbf{I}_{\mathrm{H}} \cdot \mathbf{L}) - e \mathrm{T}^{2}(\nabla \mathbf{E})\cdot \mathrm{T}^{2}(\mathbf{Q}).
\label{eq:AHamiltonian}
\end{equation}
Here $\mathbf{L}$ is the electronic orbital angular momentum, while $\mathbf{I}_{\mathrm{B}}$ and $\mathbf{I}_{\mathrm{H}}$ are the nuclear spins of the $^{11}$B and H nuclei ($I_B=3/2$ and $I_H=1/2$). The coefficients $a_\mathrm{B}$ and $a_\mathrm{H}$ determine the interaction strength between the nuclear magnetic moments and the magnetic field at the nuclei arising from the motion of the electrons. The final term in (\ref{eq:AHamiltonian}) is the interaction of the electric quadrupole moment of the boron nucleus, $Q$, with the electric field gradient at the nucleus, and it is represented as the scalar product of two second-rank tensors. The matrix elements of this electric quadrupole interaction are given by equation (8.382) of reference \cite{BNC}. Two coefficients appear in these matrix elements, $e q_{0} Q$ and $e q_{2} Q$, where $q_{0}$ represents the electric field gradient in the direction of the internuclear axis, and $q_{2}$ the field gradient in the perpendicular direction. The total electronic angular momentum is  $\mathbf{J}$. As indicated in figure~\ref{fig:levelstruct}, the $J=1$ level is split into two components of opposite parity. These are each split into three components labelled by the intermediate quantum number $F_{1}$, where $\mathbf{F}_{1}=\mathbf{I}_{\mathrm{B}}+\mathbf{J}$, each of which is further split into two levels labelled by the total angular momentum quantum number $F$. The matrix elements of the electric quadrupole interaction are proportional to $e q_{0} Q - e q_{2} Q$ for the positive parity component, and to $e q_{0} Q + e q_{2} Q$ for the negative parity component, and so these two contributions can be separated by measuring both the R(0) and Q(1) optical spectra.

\begin{figure}
	\centering
		\includegraphics[width=0.45\textwidth]{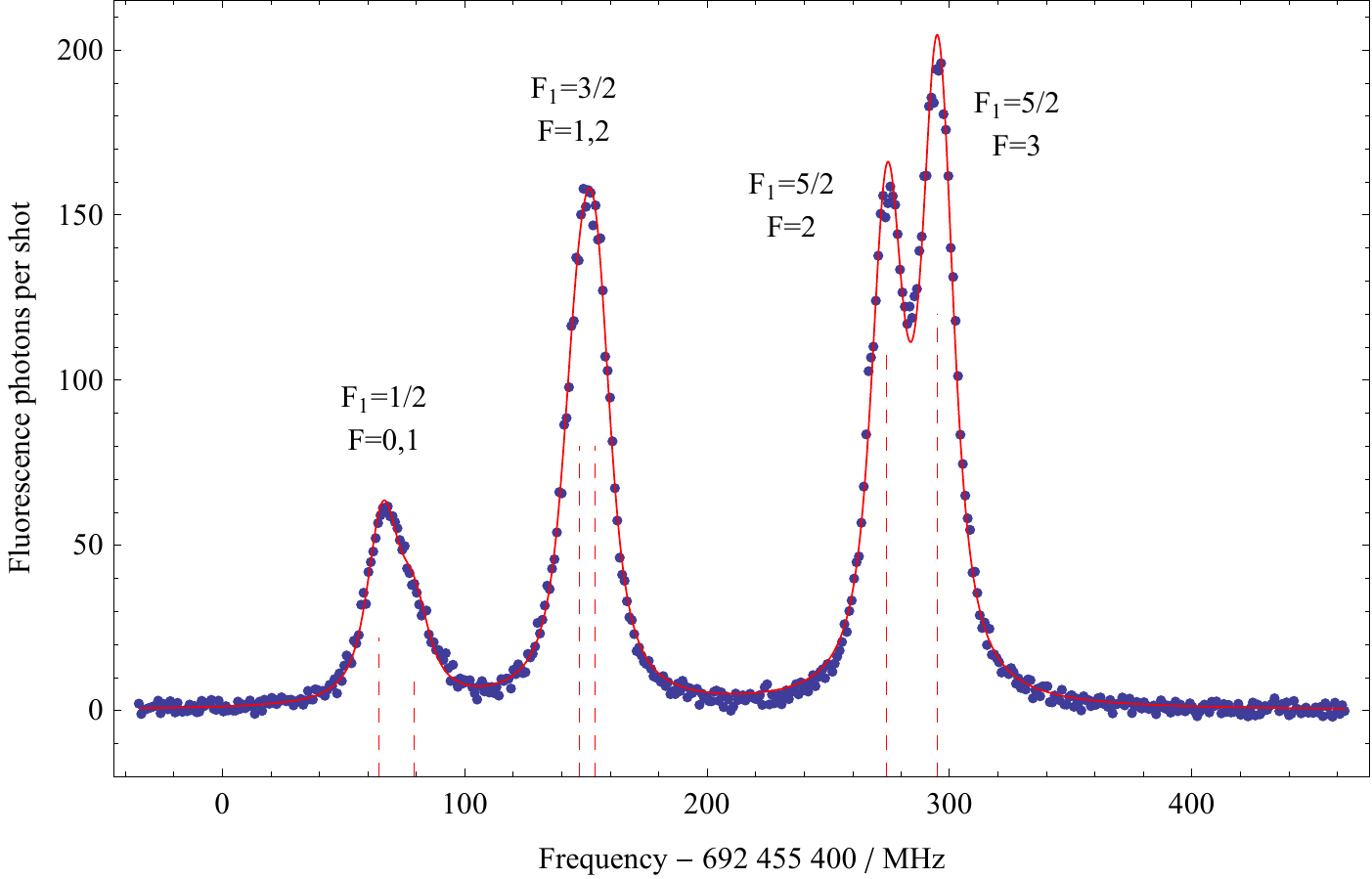}
		\caption{Optical spectrum of the R(0) transition showing the hyperfine structure in the ${\rm A}^1 \Pi (v'{=}0, J'{=}1)$ state. The solid line is a fit to a sum of six Voigt functions as discussed in the text. The dashed vertical lines show the fitted centres of each component.}
		\label{fig:optical}
\end{figure}

Figure \ref{fig:optical} shows the laser-induced fluorescence spectrum of the R(0) component of the ${\rm A}^1 \Pi (v'{=}0) \longleftarrow {\rm X}^1 \Sigma^{+} (v''{=}0)$ transition, measured by recording the sum of the signals from the two PMTs as a function of laser frequency. A similar spectrum is obtained for the Q(1) transition. As we will see below, the hyperfine structure of the lower level is tiny, and so all the structure observed in the optical spectrum is due to the hyperfine splitting of the ${\rm A}^1 \Pi (v'{=}0,J'{=}1)$ state. The three main lines in the spectrum correspond to the three $F_1$ components arising from the boron nuclear spin. The splitting due to the hydrogen nuclear spin is resolved in the $F_{1}=5/2$ component, results in a shoulder in the $F_{1}=1/2$ component, and is unresolved in $F_{1}=3/2$. To the data in figure~\ref{fig:optical} we fit a sum of six Voigt functions, with Lorentzian and Gaussian widths common to each component. The resulting fit is shown by the solid line in figure \ref{fig:optical}. For the Gaussian component, due to Doppler broadening, we find the full width at half maximum to be 11\,MHz, while for the Lorentzian component, due to power broadening, it is 14\,MHz. From these fits we obtain the hyperfine intervals with an accuracy limited by the frequency calibration of the laser scan. We analyze the Q(1) data in a similar way. We fit the measured hyperfine intervals to the eigenvalues of Hamiltonian~(\ref{eq:AHamiltonian}) to determine the hyperfine constants given in table~\ref{tab:hypconstants}. With these best fit values, the experimental and theoretical hyperfine intervals all agree within one standard deviation, showing that Hamiltonian~(\ref{eq:AHamiltonian}) is adequate to describe the data.

\begin{table}[!t]
\begin{centering}
{\begin{tabular}{ p{1.3cm}p{0.6cm}p{2.2cm}p{1.1cm} }
\toprule
& & Experimental & Theoretical \\ \hline
\multirow{6}{*}{${\rm X}^1\Sigma^{+}(v{=}0)$} & $f_{01}$ & 708309.211 (9) & --- \\
& $eq_{0}Q$ & -6.43 (5) & -6.5951 \\
& $c_{\mathrm{B}}$ & 0.410 (5) & 0.45374 \\
& $c_{\mathrm{H}}$ & $|c_{\mathrm{H}}|<$ 0.04 & -0.01370 \\
& $c_3$ & $|c_3|<$ 0.03 & 0.02020 \\
& $c_4$ & $|c_4|<$ 0.03 & --- \\
\hline
\multirow{3}{*}{${\rm A}^1\Pi(v{=}0)$} & $a_{\mathrm{B}}$ & 108 (11) & --- \\
& $a_{\mathrm{H}}$ & 36 (5) & --- \\
& $eq_{0}Q$ & 30 (6) & ---\\
& $eq_{2}Q$ & -27 (6) & ---\\
\botrule
\end{tabular}}{}
\caption{Lowest rotational transition frequency, and hyperfine constants for $^{11}$BH.  All values are in MHz.  The theoretical values are derived from \cite{Vojtik91} and \cite{Sauer95}.}
\label{tab:hypconstants}
\end{centering}
\end{table}

\subsection{Lowest rotational transition and hyperfine structure of the ground state}

To form a closed laser cooling cycle, it is necessary to address all relevant ground-state hyperfine components. If this hyperfine splitting is smaller than the natural linewidth of the transition, all components will be driven strongly using just a single laser frequency. For larger splittings, it may be necessary to apply sidebands to the laser light. The ground-state hyperfine structure of BH has not previously been measured, but calculations \cite{Vojtik91, Sauer95} suggest splittings of order 1\,MHz, roughly equal to the linewidth. The hyperfine structure is shown schematically in figure \ref{fig:levelstruct}, and is described by the Hamiltonian given in~\cite{Rothstein69}. The hyperfine structure in $J=1$ is dominated by the interaction of the electric quadrupole moment of the boron nucleus with the local electric field gradient (coefficient $- e q_{0} Q$), and the interaction of the magnetic dipole moment of the boron nucleus with the magnetic field arising from the rotation of the molecule (coefficient $c_{\mathrm{B}}$). Together, these give rise to a splitting into three main components with intermediate quantum number $F_{1}=1/2,3/2,5/2$. When the hydrogen nuclear spin is included, each of these is split into a pair of levels with total angular momentum quantum number $F$. This splitting is very small and is due to the interaction of the hydrogen magnetic dipole moment with the magnetic field of the rotating molecule (coefficient $c_{\mathrm{H}}$), along with the direct and electron-mediated interaction between the two nuclear spins (coefficients $c_3$ and $c_4$). For $J=0$, only the $c_4$ term is present and the hyperfine structure is exceedingly small.

\begin{figure}
	\centering
		\includegraphics[width=0.45\textwidth]{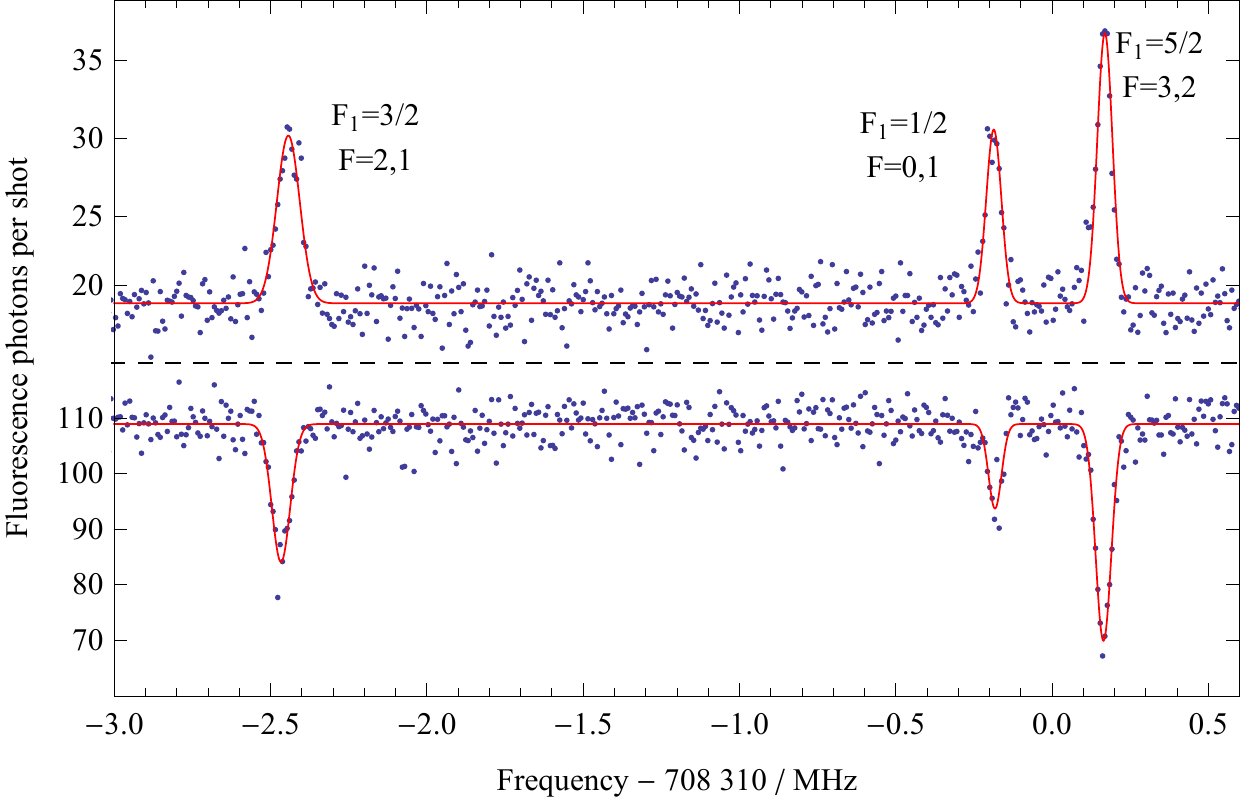}
		\caption{Millimetre-wave spectra of the lowest rotational transition of X, showing the main hyperfine structure in the $J{=}1$ state. In the lower trace the population is measured in $J=0$ as the millimetre-wave frequency is scanned, while in the upper trace the population in $J{=}1$ is measured. The solid lines are fits to a sum of three Gaussians.}
		\label{fig:THz}
\end{figure}

We measure the ground state $J=1$ hyperfine structure by driving the millimetre-wave transition between $J=0$ and $J=1$. Figure~\ref{fig:THz} shows typical millimetre-wave spectra obtained by measuring the fluorescence on either R(0) or Q(1) as a function of the millimetre-wave frequency, which is stepped between shots of the experiment in a random order. In the lower trace the probe laser is resonant with the $J{=}0$ sublevels, and we see a drop in the fluorescence rate as molecules are driven into the $J{=}1$ sublevels by the millimetre-wave radiation. With the laser instead locked to the Q(1) transition, we see a corresponding resonant increase in the fluorescence rate, as shown in the upper graph. After accounting for the molecular population already in the $J{=}1$ state, we estimate a peak transfer efficiency of about 40\%, limited by the available millimetre-wave power. In these spectra we observe the splitting into the three $F_{1}$ components arising from the boron nuclear spin. The smaller splitting due to the hydrogen nuclear spin is not resolved. We fit the spectra to a sum of three Gaussians to obtain the centre frequencies of the three components. Assignment of the lines is aided by the observation that the number of components in the spectrum depends on which of the excited state hyperfine components of the Q(1) transition is used for detection (see figure \ref{fig:optical}). When the laser is tuned to excite the $F_{1}=3/2$ component of the A state, all three hyperfine components of the rotational transition are observed, as shown in figure~\ref{fig:THz}. However, if the $F_{1}=5/2\,(1/2)$ component of the A state is used for detection, the $F_{1}=1/2\,(5/2)$ component of the rotational transition disappears.

Inevitably, the rotational spectrum is Doppler shifted because the millimetre-wave beam is not perfectly perpendicular to the molecular beam. We measure this shift by changing the speed of the molecules. This is done by mixing the diborane and argon gas with helium in various ratios. The reduction in the average mass of the carrier gas increases the speed of the molecular beam. The fitted line centres shift linearly with the speed, with a gradient of 30\,Hz(m/s)$^{-1}$. We extrapolate to zero speed to obtain the Doppler-free centre frequencies.  These frequencies are fitted to the model $f_{01} + \langle H_{\textrm{hfs}} \rangle$, where $f_{01}$ is the hyperfine-free rotational transition frequency, and $\langle H_{\textrm{hfs}} \rangle$ are the eigenvalues of the hyperfine Hamiltonian in \cite{Rothstein69}, with $c_{\mathrm{H}}$, $c_{3}$ and $c_{4}$ all set to zero. The values of $f_{01}$, $e q_{0} Q$ and $c_{\mathrm{B}}$ determined this way are given in table~\ref{tab:hypconstants}. Their uncertainties are dominated by the unresolved sub-structure due to the hydrogen nuclear spin, which limits the precision to roughly the linewidth. The two hyperfine constants agree with the theoretical predictions \cite{Vojtik91, Sauer95} to within 2.5\% and 10\% respectively. Using the measured linewidths, and the fact that all three widths are equal within their uncertainties, we obtain the upper limits to $c_{\mathrm{H}}$, $c_{3}$ and $c_{4}$ given in table~\ref{tab:hypconstants}. These are also consistent with the theoretical predictions.

%%%%%%%%%%%%%%%%%%%%%%%%%%%%%%%%%%%%%%%
%%%%%%%%%%%%%%%%%%%%%%%%%%%%%%%%%%%%%%%

\section{Discussion}

The measurements presented here indicate that BH is a good choice of molecule for laser cooling on the ${\rm A}(v'{=}0) \leftrightarrow {\rm X}(v''{=0})$ Q(1) transition. With just two lasers, one addressing the $v''=0$ branch at 433\,nm and the other re-pumping the $v''=1$ branch at 481\,nm, each molecule will scatter an average of 1000 photons before being optically pumped to a higher vibrational state. This corresponds to a change in velocity of 77\,m/s, sufficient to capture a large fraction of the velocity distribution emitted by a cryogenic buffer gas source \cite{Lu11}. The ground state hyperfine structure spans 2.7\,MHz, about twice the natural linewidth of the transition. To optimize the scattering rate it is necessary to broaden the linewidth of the cooling laser to about 3\,MHz so that all hyperfine components are addressed. This increases the total power that is needed by a factor of about 4, but this is not too problematic since the saturation intensity is only 2\,mW/cm$^{2}$. Since the laser cooling transition is from $J''=1$ to $J'=1$, there will be a dark state for any laser polarization, but this can be avoided by modulating the laser polarization \cite{Berkeland02}. With a third laser to repump molecules from $v''=2$ to $v'=1$ at 483\,nm, our measurements show that more than 10000 photons will be scattered before decaying to a higher lying vibrational state, sufficient to slow to rest a supersonic beam of BH and load it into a magneto-optical trap. To follow the changing Doppler shift as the molecules slow down, a Zeeman slower could be used, taking advantage of the strong Zeeman effect in the excited state and very weak Zeeman effect in the ground state, which greatly simplifies the required spectrum of laser light. This is an advantage of the simple structure of BH compared to the molecules that have so far been cooled. Taking into account the number of resolved ground and excited-state levels, and the polarization modulation, a scattering rate of $\Gamma/10$ is reasonable \cite{Tarbutt2013a}. Then, the stopping distance for a 430\,m/s beam is 1.5\,m. In a well-optimized magneto-optical trap, we would expect the molecules to reach the Doppler temperature of 30\,$\mu$K, or if polarization-gradient cooling is effective, the recoil temperature of 4\,$\mu$K. This gas of ultracold polar molecules would be an excellent system for exploring large-scale entanglement and the behaviour of a controlled strongly-interacting quantum system. Here, a key ingredient will be the controlled manipulation of the rotational state. In the present work we were able to drive the rotational transition at 708\,GHz with an efficiency of about 40\%, limited by the available power of less than 1\,$\mu$W and the interaction time of about 10\,$\mu$s.

\section*{Acknowledgement}
We thank Jon Dyne, Steve Maine and Valerijus Gerulis for their expert technical assistance. We are grateful to Ed Hinds for support and advice and to Ed Grant for advice on BH beam production. This work was supported by the EPSRC.

\bibliography{BH_spectroscopy}{}
\bibliographystyle{unsrt}

\end{document}